\documentclass[12pt]{article}
\usepackage{amssymb}
\usepackage{epsfig}
\usepackage{float}
\usepackage{graphicx}
\usepackage{amsmath}
\newcommand{\nua}[1]{\ensuremath{\rlap
           {\kern-2.5pt\ensuremath
           {\overset{\scriptscriptstyle(-)}{\phantom{\nu}}}}
           {\ensuremath{{\nu}_{#1}}}}}
\begin{document}
\begin{center}
{\bf Basics of General Theory of Relativity for Beginners}
\end{center}
\begin{center}
S. M. Bilenky
\end{center}
\begin{center}
{\em  Joint Institute for Nuclear Research, Dubna, R-141980,
Russia\\}
{\em TRIUMF
4004 Wesbrook Mall,Vancouver BC, V6T 2A3
Canada\\}
\end{center}

\begin{abstract}
We present a basics of the Einstein General Theory of Relativity. In the first part of this review we derive relations of Riemann geometry which are used in the General Relativity. In the second part  we discuss Einstein Equations and some of its consequences (The Schwarzschild solution, gravitational waves, Friedman Equations etc). In the Appendix we
 briefly discuss a history of the discovery of the Einstein Equations.
\end{abstract}

\section{Introduction}
General Theory of Relativity is a great  theory, confirmed by all existing data (see, for example, "Experimental Tests of Gravitational Theory" by  T. Damour in PDG \cite{Zyla:2020zbs}) It is based on the requirement of invariance under the general transformations of coordinates in a curved Riemann space. 
The General Theory of Relativity, apparently, support a suggestion\footnote{"Simplicity is a guide to the theory choice" A. Einstein.}  that in a correct theory  the simplest possibilities are realized. Recent discovery of the predicted by GTO gravitational waves opened a new and very powerful way of the investigation of the Universe. In the book "Classical Theory of Fields" L.D. Landau and E.M. Lifshitz wrote : "The General Theory of Relativity which was created by Einstein (and finally formulated by him in 1916) is, apparently, the most beautiful of all existing physical theories. It is remarkable that it was built by Einstein in a purely deductive way and only later was confirmed by astronomical observations" In this review I tried to present the basics of the General Theory of Relativity and some of its consequences in such a way that all derivations can be easily followed by a reader.

\section{Curved Riemann space. Vectors, Tensors}
The   General Theory of Relativity (GTR)
 is based  on the requirement of the invariance of basic equations under transformations of the coordinates in  a general coordinate system\footnote{For textbooks  see, for example, \cite{Dirac,Landau,Zeldovich}.} In order to formulate the  equations of the General Theory of Relativity A. Einstein used mathematical methods of the non Euclidian geometry which we briefly consider now.

Let $x^{\alpha}$ ($\alpha=0,1,2,3$) be coordinates of a point in a general coordinate system. The square of the length between
infinitesimally close
points $x^{\alpha}+dx^{\alpha}$ and $x^{\alpha}$ (interval) has the form
\begin{equation}\label{metric}
ds^{2}=g_{\alpha\beta}(x)~dx^{\alpha}dx^{\beta}.
\end{equation}
It is obvious that $g_{\alpha\beta}(x)=g_{\beta\alpha}(x)$. $g_{\alpha\beta}(x)$ is called the metric tensor.

Let $x'$ and $x$ be coordinates of a point in different reference systems. The coordinates  $x'$ are functions of $x$ (and vice versa). Thus, we have
\begin{equation}\label{transf1}
d x'^{\alpha}=\frac{\partial x'^{\alpha}}{\partial x^{\rho}}d x^{\rho}.
\end{equation}
If we multiply (\ref{transf1}) by $\frac{\partial x^{\sigma}}{\partial x'^{\alpha}}$ and sum over $\alpha$ we find the inverse transformation
\begin{equation}\label{transf2}
d x^{\sigma}=\frac{\partial x^{\sigma}}{\partial x'^{\alpha}}d x'^{\alpha}.
\end{equation}
 {\em Contravariant vector  $A^{\alpha}(x)$}   transforms as the differential of coordinates:
\begin{equation}\label{Transf2}
A'^{\alpha}(x')=\frac{\partial x'^{\alpha}}{\partial x^{\rho}} A^{\rho}(x).
\end{equation}
 Interval $ds^{2}$ is an invariant. From (\ref{metric}) and (\ref{transf1}) we find
\begin{equation}\label{metric2}
ds^{2}=g_{\alpha\beta}(x)~dx^{\alpha}dx^{\beta}=g_{\alpha\beta}(x)~\frac{\partial x^{\alpha}}{\partial x'^{\rho}}\frac{\partial x^{\beta}}{\partial x'^{\sigma}}~dx'^{\rho}dx'^{\sigma}=
g'_{\rho\sigma}(x')~dx'^{\rho}dx'^{\sigma}.
\end{equation}
Thus, the metric tensor $g_{\alpha\beta}$ is transformed as follows
\begin{equation}\label{metric3}
g'_{\rho\sigma}(x')=\frac{\partial x^{\alpha}}{\partial x'^{\rho}}\frac{\partial x^{\beta}}{\partial x'^{\sigma}}~g_{\alpha\beta}(x).
\end{equation}
From (\ref{metric3}) we find the inverse transformation
\begin{equation}\label{metric4}
g_{\alpha\beta}(x)=\frac{\partial x'^{\rho}}{\partial x^{\alpha}}\frac{\partial x'^{\sigma}}{\partial x^{\alpha}}~g'_{\rho\sigma}(x')
\end{equation}
The scalar product of two vectors $A^{\alpha}$ and $B^{\alpha}$ is determined as follows
\begin{equation}\label{scalar}
    A(x)\cdot B(x)=g_{\alpha\beta}(x)A^{\alpha}(x)B^{\beta}(x).
\end{equation}
The scalar product  is an invariant. In fact, taking into account (\ref{metric3}) and (\ref{Transf2}), we have
\begin{equation}\label{transf3}
A(x)\cdot B(x)=g_{\alpha\beta}(x)\frac{\partial x^{\alpha}}{\partial x'^{\rho}}\frac{\partial x^{\beta}}{\partial x'^{\sigma}}A'^{\rho}(x) B'^{\sigma}(x)=g'_{\rho\sigma}(x')A'^{\rho}(x') B'^{\sigma}(x')=A'(x')\cdot  B'(x').
\end{equation}
{\em Covariant vector $A_{\alpha}$} is determined by the relation
\begin{equation}\label{transf4}
A_{\alpha}=g_{\alpha\beta}~A^{\beta}.
\end{equation}
Thus, the scalar product has the form
\begin{equation}\label{tran4}
 A\cdot B= A_{\alpha} B^{\alpha}.
\end{equation}
Further, we have
\begin{equation}\label{transf15}
 A_{\alpha}(x)B^{\alpha}(x)=A_{\alpha}(x)\frac{\partial x^{\alpha}}{\partial x'^{\rho}}B'^{\rho}(x')=
A'_{\rho}(x')B'^{\rho}(x').
\end{equation}
From this relation we find that a covariant vector is transformed in the following way
\begin{equation}\label{transf16}
 A'_{\rho}(x')=\frac{\partial x^{\alpha}}{\partial x'^{\rho}}A_{\alpha}(x).
\end{equation}
From (\ref{transf16}) we find the inverse transformation
\begin{equation}\label{transf17}
 A_{\beta}(x)=\frac{\partial x'^{\rho}}{\partial x^{\beta}}A'_{\rho}(x').
\end{equation}
Tensors are transformed as product of vectors. For example, the second rank tensors $T^{\rho\sigma}$ and
$T^{\rho}_{\sigma}$ are transformed , respectively, as product of two contravariant vectors and product of
contravariant and covariant vectors:
\begin{equation}\label{transf18}
T'^{\alpha\beta}(x')
=\frac{\partial x'^{\alpha}}{\partial x^{\rho}}\frac{\partial x'^{\beta
}}{\partial x^{\sigma}}
T^{\rho\sigma}(x),~~~T'^{\alpha}_{\beta}(x')
=\frac{\partial x'^{\alpha}}{\partial x^{\rho}}
\frac{\partial x^{\sigma}}{\partial x'^{\beta}}T^{\rho}_{\sigma}(x).
\end{equation}
The tensor of the third rank $T^{\rho\sigma}_{\nu}$ is transformed as the product of two contravariant vectors and  covariant vector:
\begin{equation}\label{transf19}
T'^{\alpha\beta}_{\mu}(x')=
\frac{\partial x'^{\alpha}}{\partial x^{\rho}}\frac{\partial x'^{\beta}}{\partial x^{\sigma}}\frac{\partial x^{\nu}}{\partial x'^{\mu}}T^{\rho\sigma}_{\nu}(x),
\end{equation}
etc. From (\ref{metric3}) it follows that $g_{\alpha\beta}$ is the tensor of the second rank.

Let us put in (\ref{transf19}) $\beta=\mu$ and sum up over $\mu$. We have                                                                                                                                           \begin{equation}\label{transf20}
T'^{\alpha\mu}_{\mu}(x')=
\frac{\partial x'^{\alpha}}{\partial x^{\rho}}\delta^{\nu}_{\sigma}T^{\rho\sigma}_{\nu}(x)=
\frac{\partial x'^{\alpha}}{\partial x^{\rho}}T^{\rho\sigma}_{\sigma}(x).
\end{equation}
Thus, $T^{\rho\sigma}_{\sigma}$ is a vector.

The metric tensor $g^{\rho\alpha}=g^{\alpha\rho}$ is determined as follows
\begin{equation}\label{transf11}
g^{\rho\alpha}g_{\alpha\beta}=\delta^{\rho}_{\beta}.
\end{equation}
From (\ref{transf4}) and (\ref{transf11}) we find
\begin{equation}\label{transf13}
  A^{\rho}=g^{\rho\alpha}~A_{\alpha}.
\end{equation}
Thus, with the help of the tensor $g_{\rho\alpha}$ ($g^{\rho\alpha}$) we can lower (raise) indexes.  For the scalar product  we  have
\begin{equation}\label{transf14}
A\cdot B=g_{\alpha\beta}g^{\alpha\rho}A_{\rho}g^{\beta\sigma}B_{\sigma}=
g^{\rho\sigma}A_{\rho}B_{\sigma}.
\end{equation}
In the flat Minkowski  space the derivative of a vector is a tensor. Let us consider the derivative of a vector $ A^{\alpha}$. From (\ref{Transf2}) we find
\begin{equation}\label{transf15}
\frac{\partial A'^{\alpha}(x')}{\partial x'^{\beta}}=\frac{\partial x'^{\alpha}}{\partial x^{\rho}}\frac{\partial x^{\sigma}}{\partial x'^{\beta}}\frac{\partial A^{\rho}(x)}{\partial x^{\sigma}}+
\frac{\partial^{2} x'^{\alpha}}{\partial x^{\rho}\partial x^{\sigma}}\frac{\partial x^{\sigma}}{\partial x'^{\beta}}A^{\rho}(x)
\end{equation}
Thus, in a  curved Riemann space the derivative of the vector is not a tensor. This is connected with the fact  that at different points $x$ and $x+dx$ vector $A^{\alpha}$ is transformed differently.

In the next section we will introduce a notion of {\em a covariant derivative}. The covariant derivative will be introduced in such a way that the  derivative of a vector is a tensor, derivative of a tensor is a tensor of higher rank etc.  Covariant derivatives  allow to formulate invariant under general transformations of coordinates basic equations of the General Theory of Relativity.

\section{Parallel Displacement of Vectors. Christoffel Symbols. Covariant Derivative}
In order to introduce the covariant derivative we need first to define the operation of parallel displacement of a vector
in a curved space from a point $x$ to the point $x+dx$. For that let us imbed a four-dimensional curved space in a flat space of a dimension $N>4$. Coordinates of a point in this space  will be denoted $z^{n}$. The square of distance between points $z^{n}+dz^{n}$ and $z^{n}$ is given by
\begin{equation}\label{Ndim}
    ds^{2}=h_{nm}dz^{n}dz^{m},
\end{equation}
where the metric tensor $h_{nm}$ is a constant. We can write down $ds^{2}$ also in the form
\begin{equation}\label{Ndim1}
 ds^{2} =dz^{n}dz_{n},
\end{equation}
where  $dz_{n}=h_{nm}dz^{m}$.

Physical space is a ``surface" in the N-dimensional space. To each point $x$ of the surface corresponds a point $y^{n}(x)$ in $N$-dimensional flat space. We have
\begin{equation}\label{Ndim2}
    \delta y^{n}=y_{,\alpha}^{n}~\delta x^{\alpha},
\end{equation}
where the following  notation
\begin{equation}\label{Ndim3}
    \frac{\partial y^{n}}{\partial x^{\alpha}}=y_{,\alpha}^{n}
\end{equation}
is used. The square of the distance between points $x+dx$ and $x$ on the surface is given by
\begin{equation}\label{Ndim4}
\delta s^{2}=h_{nm}\delta y^{n}\delta y^{m}=h_{nm}y_{,\alpha}^{n}y_{,\beta}^{m}\delta x^{\alpha}\delta x^{\beta} =y_{,\alpha}^{n} y_{n,\beta}\delta x^{\alpha}\delta x^{\beta}.
\end{equation}
Notice, that we take into account in (\ref{Ndim4}) that $h_{nm}$ is a constant. The equation (\ref{Ndim4}) has the standard form
\begin{equation}\label{Ndim5}
 \delta s^{2}=g_{\alpha\beta}\delta x^{\alpha}\delta x^{\beta},   \end{equation}
where the metric tensor $g_{\alpha\beta}$ is given by
\begin{equation}\label{Ndim5a}
g_{\alpha\beta}=y_{,\alpha}^{n} y_{n,\beta}.
\end{equation}
Let us consider a vector $A^{\alpha}(x)$ in the four-dimensional space. From (\ref{Ndim2}) follows that the corresponding vector in the N-dimensional space is given by the relation
\begin{equation}\label{Ndim5c}
 A^{n}(x)=y^{n}_{,\alpha}(x)A^{\alpha}(x).
\end{equation}
Let us perform the parallel displacement of the vector $ A^{n}$  from the point $y^{n}(x)$ to the point $y^{n}(x+dx)$. The transferred vector in general will not  lie down on the surface. We can present it in the form
\begin{equation}\label{Ndim6}
A^{n}=A^{n}_{\mathrm{tan}}+A^{n}_{\mathrm{nor}},
\end{equation}
where $A^{n}_{\mathrm{tan}}$ is the projection of $A^{n}$ on the surface and $A^{n}_{\mathrm{nor}}$ is orthogonal to  $A^{n}_{\mathrm{tan}}$ vector:
\begin{equation}\label{Ndim6a}
 A^{n}_{\mathrm{tan}}(A_{n})_{\mathrm{nor}}=0.
\end{equation}
If a vector $K^{\alpha}(x+dx)$ on the surface corresponds to the vector $A^{n}_{\mathrm{tan}}$ we have
\begin{equation}\label{Ndim7}
A^{n}_{\mathrm{tan}}=y^{n}_{,\alpha}(x+dx)K^{\alpha}(x+dx).
\end{equation}
Let us now multiply (\ref{Ndim7}) by $y_{n,\beta}(x+dx)$. Using (\ref{Ndim5a}), we find
\begin{equation}\label{Ndim7a}
    g_{\alpha\beta}(x+dx)K^{\alpha}(x+dx)=K_{\beta}(x+dx)=
    A^{n}_{\mathrm{tan}}y_{n,\beta}(x+dx).
\end{equation}
From (\ref{Ndim6a}) and (\ref{Ndim7}) we find that
\begin{equation}\label{Nd}
A^{n}_{\mathrm{nor}}y_{n,\beta}(x+dx)=0.
\end{equation}
Taking into account this relation, from (\ref{Ndim7a}) we have\footnote{Notice that we took into account that
components of a vector  in the flat N-dimensional space are not changed under the parallel displacement.}
\begin{eqnarray}\label{Ndim7b}
K_{\beta}(x+dx)&=&A^{n}y_{n,\beta}(x+dx)
=A^{\alpha}(x)y^{n}_{,\alpha}(x) (y_{n,\beta}(x)+ y_{n,\beta\sigma}(x)dx^{\sigma})\nonumber\\&=&A_{\beta}(x)+\delta A_{\beta}(x)
\end{eqnarray}
Here
\begin{equation}\label{Ndim12}
\delta A_{\beta}=A^{\alpha}(x)y^{n}_{,\alpha}(x)y_{n,\beta\sigma}(x)dx^{\sigma}.    \end{equation}
Thus, $\delta A_{\beta}$ is a change of the covariant vector $A_{\beta}$ under the parallel displacement.

The term $y^{n}_{\alpha}y_{n,\beta \sigma}$  can be expressed through derivatives of the metric tensor $g_{\alpha\beta}$. In fact, from (\ref{Ndim5a}) we have
\begin{equation}\label{Cristo2}
g_{\alpha\beta,\sigma}=y^{n}_{,\alpha\sigma}y_{n,\beta}+
y^{n}_{,\alpha}y_{n,\beta\sigma}.
\end{equation}
Let us  change   $\beta\leftrightarrows\sigma$ in (\ref{Cristo2}). We find the relation
\begin{equation}\label{Cristo3}
g_{\alpha\sigma,\beta}=y^{n}_{,\alpha\beta}y_{n,\sigma}+
y^{n}_{,\alpha}y_{n,\sigma\beta}.
\end{equation}
It is obvious that $y_{n,\beta\sigma}=y_{n,\sigma\beta}$.
Thus, the last two terms in (\ref{Cristo2}) and (\ref{Cristo3}) are equal. Let us now change in (\ref{Cristo2}) $\alpha\leftrightarrows\sigma$. We find the relation
\begin{equation}\label{Cristo4}
g_{\sigma\beta,\alpha}=y^{n}_{,\sigma\alpha}y_{n,\beta}+
y^{n}_{,\sigma}y_{n,\beta\alpha}
\end{equation}
It is obvious that this term is equal to the sum of first terms of (\ref{Cristo2}) and (\ref{Cristo3}). Thus,  we obtain the following relation
\begin{equation}\label{Cristo4a}
g_{\alpha\beta,\sigma}+g_{\alpha\sigma,\beta}=g_{\sigma\beta,\alpha} +2 y^{n}_{,\alpha}y_{n,\beta\sigma}.
\end{equation}
Finally we have
\begin{equation}\label{Cristo5}
y^{n}_{\alpha}y_{n,\beta \sigma}=\frac{1}{2}(g_{\alpha\beta,\sigma}+g_{\alpha\sigma,\beta}-
g_{\sigma\beta,\alpha}).
\end{equation}
The right-hand part of this equation is  called the Christoffel symbol:
\begin{equation}\label{Cristo6}
\Gamma_{\alpha\beta\sigma}=
\frac{1}{2}(g_{\alpha\beta,\sigma}+
g_{\alpha\sigma,\beta}-g_{\sigma\beta,\alpha})
\end{equation}
From (\ref{Ndim12}) and (\ref{Cristo5}) we have
\begin{equation}\label{Cristo7}
 \delta A_{\beta}= A^{\alpha}\Gamma_{\alpha\beta\sigma}dx^{\sigma}.
\end{equation}
It is obvious that  $\delta A_{\beta}$ can be also presented in the form
\begin{equation}\label{Cristo7a}
 \delta A_{\beta}= \Gamma^{\alpha}_{\beta\sigma}A_{\alpha}dx^{\sigma},
\end{equation}
where
\begin{equation}\label{Cristo7b}
\Gamma^{\alpha}_{\beta\sigma}=g^{\alpha\rho}\Gamma_{\rho\beta\sigma}.    \end{equation}
We conclude from (\ref{Cristo6}) and (\ref{Cristo7}) that the parallel displacement of a vector in four-dimensional non Euclidian space is determined by derivatives of the metric tensor
$g_{\alpha\beta}$ (there are no more  references to the N-dimensional flat space).

It is easy to see that under the change $\alpha\leftrightarrows\beta$ the first term in the right-hand  side of the Christoffel symbol (\ref{Cristo6}) is symmetric and the last two terms are antisymmetric. Thus, we have the relation
\begin{equation}\label{Cristo8}
\Gamma_{\alpha\beta\sigma}+\Gamma_{\beta\alpha\sigma}=
g_{\alpha\beta,\sigma}.
\end{equation}
It is  also obvious that
\begin{equation}\label{Cristo9}
\Gamma_{\alpha\beta\sigma}=\Gamma_{\alpha\sigma\beta}.
\end{equation}
Under the parallel displacement the length of a vector is not changed. In fact, we have
\begin{eqnarray}\label{Cristo10}
\delta(g^{\alpha\beta}A_{\alpha}A_{\beta})&=&g^{\alpha\beta}\delta A_{\alpha}
A_{\beta}+g^{\alpha\beta}A_{\alpha}\delta A_{\beta}
+\delta g^{\alpha\beta}A_{\alpha}A_{\beta}\nonumber\\&=&
[A^{\rho}A^{\alpha}
(\Gamma_{\rho\alpha\sigma}+ \Gamma_{\alpha\rho\sigma} )
+g^{\alpha\beta}_{,\sigma}A_{\alpha}A_{\beta}]dx^{\sigma}
\end{eqnarray}
Further, using relation (\ref{Cristo8}) we find
\begin{equation}\label{Cristo11}
\delta(g^{\alpha\beta}A_{\alpha}A_{\beta})=
[g_{\alpha\rho,\sigma}A^{\alpha}A^{\rho}
+g^{\alpha\beta}_{,\sigma}A_{\alpha}A_{\beta}]dx^{\sigma}.
\end{equation}
We have
\begin{equation}\label{Cristo12}
g^{\alpha\beta}g_{\alpha\rho}=\delta^{\beta}_{\rho}.
\end{equation}
From this relation we find
\begin{equation}\label{Cristo13}
g^{\alpha\beta}_{,\sigma}g_{\alpha\rho}+
g^{\alpha\beta}g_{\alpha\rho,\sigma}=0.
\end{equation}
If we multiply this relation by $g_{\beta\tau}$, we obtain the following relation
\begin{equation}\label{Cristo14}
g_{\tau\rho,\sigma}=-
g^{\alpha\beta}_{,\sigma}g_{\alpha\rho}g_{\beta\tau}
\end{equation}
From (\ref{Cristo14}) we have
\begin{equation}\label{Cristo15}
g_{\tau\rho,\sigma}A^{\tau}A^{\rho}
=-g^{\alpha\beta}_{,\sigma}g_{\alpha\rho}g_{\beta\tau}A^{\rho} A^{\sigma}
=-g^{\alpha\beta}_{,\sigma}A_{\alpha}A_{\beta}.
\end{equation}
Thus, finally, we find
\begin{equation}\label{Cristo15}
\delta(g^{\alpha\beta}A_{\alpha}A_{\beta})=0.
\end{equation}
Let us consider the vector $A+\lambda B$, where $\lambda$ is an  arbitrary constant. From the relation
\begin{equation}\label{Cristo16}
    \delta(A_{\alpha}+\lambda B_{\alpha})(A^{\alpha}+\lambda B^{\alpha})=0
\end{equation}
we have
\begin{equation}\label{Cristo17}
\delta(A_{\alpha}B^{\alpha})=0.
\end{equation}
Thus, the scalar product $(A_{\alpha}B^{\alpha})$ is not changed under the parallel displacement of vectors $A$ and $B$.

From (\ref{Cristo17}) we find
\begin{equation}\label{CCristo17}
 \delta A_{\alpha}B^{\alpha}+A_{\alpha}\delta B^{\alpha}=0
\end{equation}
Further, using (\ref{Cristo7}) from (\ref{CCristo17}) we obtain the following relation
\begin{equation}\label{Cristo18}
A_{\rho}\delta B^{\rho}=
-B^{\alpha}\Gamma_{\beta\alpha\sigma}dx^{\sigma}A^{\beta}
=-B^{\alpha}\Gamma^{\rho}_{\alpha\sigma}dx^{\sigma}A_{\rho}
\end{equation}
Thus,  for the contravariant vector  we have
\begin{equation}\label{Cristo20}
 \delta B^{\rho}=-\Gamma^{\rho}_{\alpha\sigma}B^{\alpha}dx^{\sigma}.
\end{equation}
Let us consider now the difference of the vector $A_{\alpha}(x+dx)$ and the vector  which is obtained by the parallel displacement of the vector $A_{\alpha}(x)$ from the point $x$ to the point $x+dx$. Taking into account (\ref{Cristo18}) we have
\begin{equation}\label{covar1}
A_{\alpha}(x+dx)-[A_{\alpha}(x)+\Gamma^{\beta}_{\alpha\sigma}(x)
A_{\beta}(x)dx^{\sigma}]=
[A_{\alpha,\sigma}(x)-\Gamma^{\beta}_{\alpha\sigma}(x)A_{\beta}(x)]~dx^{\sigma}.
\end{equation}
  The covariant derivative of the  vector $A_{\alpha}$ is denoted by $A_{\alpha:\sigma}$. From (\ref{covar1}) follows that it is given by the relation
\begin{equation}\label{covar2}
A_{\alpha:\sigma}=
A_{\alpha,\sigma}-\Gamma^{\beta}_{\alpha\sigma}A_{\beta}
\end{equation}
The covariant derivative  $A_{\alpha:\sigma}$ is a tensor. This follows from the fact that the left-hand side of (\ref{covar1}) is a vector.

From (\ref{Cristo20}) follows that the covariant derivative of the contravariant vector is given by
\begin{equation}\label{covar2a}
A^{\alpha}_{:\sigma}=A^{\alpha}_{,\sigma}+
\Gamma^{\alpha}_{\beta\sigma}A^{\beta}.
\end{equation}
For the covariant derivative of a product of two vectors we obviously find
\begin{eqnarray}\label{covar3}
(A_{\alpha}B_{\beta})_{:\sigma}&=&A_{\alpha:\sigma}
B_{\beta}+A_{\alpha}B_{\beta:\sigma} =
(A_{\alpha,\sigma}-\Gamma^{\rho}_{\alpha\sigma}A_{\rho})B_{\beta}
+A_{\alpha}
(B_{\beta,\sigma}-\Gamma^{\rho}_{\beta\sigma}B_{\rho})\nonumber\\&=
&(A_{\alpha}B_{\beta})_{,\sigma}-
\Gamma^{\rho}_{\alpha\sigma}A_{\rho}B_{\beta}-\Gamma^{\rho}_{\beta\sigma}A_{\alpha}B_{\rho}.
\end{eqnarray}
 Taking into account that a product of two vectors is transformed as a tensor of the second rank,  for the covariant derivative of a second rank tensor
$T_{\alpha\beta}$  we have
\begin{equation}\label{covar3a}
T_{\alpha\beta:\sigma}= T_{\alpha\beta,\sigma} -\Gamma^{\rho}_{\alpha\sigma} T_{\rho\beta} -
\Gamma^{\rho}_{\beta\sigma}T_{\alpha\rho}.
\end{equation}
It is obvious that the covariant derivative of a scalar $Y(x)$ is equal to the normal derivative:
\begin{equation}\label{covar5}
Y_{:\sigma} = Y_{,\sigma}.
\end{equation}
Finally, let us calculate the covariant derivative of the metric tensor. Using (\ref{covar3a}),  we have
\begin{equation}\label{covar6}
g_{\alpha\beta:\sigma}=g_{\alpha\beta,\sigma}
-\Gamma_{\beta\alpha\sigma}-\Gamma_{\alpha\beta\sigma}.
\end{equation}
Taking into account (\ref{Cristo8}), from this relation we find
\begin{equation}\label{covar7}
 g_{\alpha\beta:\sigma}=0.
\end{equation}

\section{The Riemann Curvature Tensor}

Let us consider a second covariant derivative of a vector $A_{\alpha:\beta:\rho}$. In a flat space the order of differentiation is not important. This is not the case in a curved Riemann  space. In fact, we have
\begin{eqnarray}\label{curv4}
A_{\alpha:\beta:\rho}&=&(A_{\alpha:\beta})_{:\rho}=A_{\alpha,\beta\rho}-\Gamma^{\sigma}_{\alpha\beta,\rho}A_{\sigma}
-\Gamma^{\sigma}_{\beta\rho}A_{\sigma,\rho}-\Gamma^{\sigma}_{\alpha\rho}A_{\sigma,\beta}
\nonumber\\&+&\Gamma^{\tau}_{\alpha\rho}\Gamma^{\sigma}_{\tau\beta}A_{\sigma}-
\Gamma^{\sigma}_{\beta\rho}A_{\alpha,\sigma}+\Gamma^{\sigma}_{\beta\rho}\Gamma^{\tau}_{\alpha\sigma}
A_{\tau}.
\end{eqnarray}
From this relation we find
\begin{equation}\label{curv6}
A_{\alpha:\beta:\rho}-A_{\alpha:\rho:\beta}=
(\Gamma^{\sigma}_{\alpha\rho,\beta}-\Gamma^{\sigma}_{\alpha\beta,\rho}
+\Gamma^{\tau}_{\alpha\rho}\Gamma^{\sigma}_{\tau\beta}-
\Gamma^{\tau}_{\alpha\beta}\Gamma^{\sigma}_{\tau\rho})A_{\sigma}.
\end{equation}
The difference $A_{\alpha:\beta:\rho}-A_{\alpha:\rho:\beta}$ characterizes the curvature of a Riemann space. The tensor
 of the fourth rank
\begin{equation}\label{curv7}
R^{\sigma}_{\alpha\beta\rho}=
\Gamma^{\sigma}_{\alpha\rho,\beta}-\Gamma^{\sigma}_{\alpha\beta,\rho}
+\Gamma^{\tau}_{\alpha\rho}\Gamma^{\sigma}_{\tau\beta}-
\Gamma^{\tau}_{\alpha\beta}\Gamma^{\sigma}_{\tau\rho}
\end{equation}
is called   {\em the Riemann curvature tensor}.\footnote{Notice that it is possible  to show that if the curvature tensor is equal to zero the space is flat. Thus, if the curvature tensor is different from zero, a four-dimensional space is  curved.}

From (\ref{curv6}) It is obvious from (\ref{curv6}) and (\ref{curv6}) that
\begin{equation}\label{curv8}
 R^{\sigma}_{\alpha\beta\rho}=-R^{\sigma}_{\alpha\rho\beta}.
\end{equation}
Further, taking into account  the relation $\Gamma^{\sigma}_{\alpha\rho}=\Gamma^{\sigma}_{\rho\alpha}$
(see (\ref{Cristo9})) it is easy to check that the Riemann curvature tensor satisfy the following cyclic relation
\begin{equation}\label{curv9}
 R^{\sigma}_{\alpha\beta\rho}+ R^{\sigma}_{\beta\rho\alpha}+  R^{\sigma}_{\rho\alpha\beta} =0.
\end{equation}

We have
\begin{equation}\label{curv10}
 R_{\sigma\alpha\beta\rho}=g_{\sigma\tau}R^{\tau}_{\alpha\beta\rho}.
\end{equation}
From (\ref{curv7}) and (\ref{curv10}) we find
\begin{equation}\label{curv11}
 R_{\sigma\alpha\beta\rho}=g_{\sigma\tau}\Gamma^{\tau}_{\alpha\rho,\beta}+
 \Gamma^{\mu}_{\alpha\rho}\Gamma_{\sigma\mu\beta}-(\rho\leftrightarrows\beta).
\end{equation}
Using (\ref{Cristo8}), for the first term in the right-hand side of (\ref{curv11}) we obviously have
\begin{equation}\label{curv12}
g_{\sigma\tau}\Gamma^{\tau}_{\alpha\rho,\beta}= \Gamma_{\sigma\alpha\rho,\beta} - g_{\sigma\tau,\beta}\Gamma^{\tau}_{\alpha\rho}=
\Gamma_{\sigma\alpha\rho,\beta}-(\Gamma_{\sigma\tau\beta}+
\Gamma_{\tau\sigma\beta})\Gamma^{\tau}_{\alpha\rho}.
\end{equation}
From (\ref{curv11}) and (\ref{curv12}) we find
\begin{equation}\label{curv13}
 R_{\sigma\alpha\beta\rho}=\Gamma_{\sigma\alpha\rho,\beta}- \Gamma^{\mu}_{\alpha\rho}\Gamma_{\mu\sigma\beta}
-(\rho\leftrightarrows\beta).
\end{equation}
Further, using  (\ref{Cristo6}) and (\ref{curv13}) we find the following expression for the Riemann curvature tensor
\begin{equation}\label{curv14}
 R_{\sigma\alpha\beta\rho}=\frac{1}{2}(g_{\sigma\rho,\alpha\beta} -g_{\rho\alpha,\sigma\beta} -
 g_{\beta\sigma,\alpha\rho}+g_{\beta\alpha,\sigma\rho})-
 \Gamma^{\mu}_{\alpha\rho}\Gamma_{\mu\sigma\beta}+
 \Gamma^{\mu}_{\alpha\beta}\Gamma_{\mu\sigma\rho}.
\end{equation}
From (\ref{curv14}) follows that the tensor $R_{\sigma\alpha\beta\rho}$ is antisymmetric under the exchange of indexes $\sigma\leftrightarrows\alpha$ or   $\beta\leftrightarrows\rho$ and is symmetric under the exchange of pairs of indexes $(\sigma\alpha)\leftrightarrows(\beta\rho)$:
\begin{equation}\label{curv15}
R_{\sigma\alpha\beta\rho}=-R_{\alpha\sigma\beta\rho},~~~
R_{\sigma\alpha\beta\rho}=-R_{\sigma\alpha\rho\beta},~~~
R_{\sigma\alpha\beta\rho}=R_{\beta\rho\sigma\alpha}.
\end{equation}

\section{Bianci identity}

The Bianci identity  play  an important role in the General Relativity.  In this section we will derive this identity.

Let us consider the second covariant derivative of  product of two vectors. We have
\begin{equation}\label{Bainc1}
 (A_{\alpha}B_{\beta})_{:\rho:\sigma}=A_{\alpha:\rho:\sigma}B_{\beta}+A_{\alpha:\rho}B_{\beta:\sigma} +
A_{\alpha:\sigma}B_{\beta:\rho}+A_{\alpha}B_{\beta:\rho:\sigma}.
\end{equation}
From this relation, taking into account (\ref{curv6}) (\ref{curv7}), we find
\begin{equation}\label{Bainc2}
  (A_{\alpha}B_{\beta})_{:\rho:\sigma}- (A_{\alpha}B_{\beta})_{:\sigma:\rho} =R^{\tau}_{\alpha\rho\sigma}A_{\tau}B_{\beta}+R^{\tau}_{\beta\rho\sigma}A_{\alpha}B_{\tau}
\end{equation}
The product $A_{\alpha}B_{\beta}$ is transferred as a tensor of a second rank. For any second rank tensor  we have the following relation
\begin{equation}\label{Bainc3}
(T_{\alpha\beta})_{:\rho:\sigma}-(T_{\alpha\beta})_{:\sigma:\rho}=R^{\tau}_{\alpha\rho\sigma}
T_{\tau\beta}+R^{\tau}_{\beta\rho\sigma}T_{\alpha\tau}.
\end{equation}
In particular,  for the tensor $A_{\alpha:\beta}$ we find
\begin{equation}\label{Bainc4}
A_{\alpha:\beta:\rho:\sigma} - A_{\alpha:\beta:\sigma:\rho}=R^{\tau}_{\alpha\rho\sigma} A_{\tau:\beta} +
R^{\tau}_{\beta\rho\sigma} A_{\alpha:\tau}.
\end{equation}
Let us perform  in (\ref{Bainc4}) the cyclic permutation of the indexes $\beta\rho\sigma$. We obtain two additional relations. The sum of all three relations is given by
\begin{eqnarray}\label{Bainc5}
&&(A_{\alpha:\beta:\rho:\sigma} - A_{\alpha:\beta:\sigma:\rho})
+(A_{\alpha:\sigma:\beta:\rho} - A_{\alpha:\rho:\beta:\sigma})+
(A_{\alpha:\rho:\sigma:\beta} - A_{\alpha:\sigma:\rho:\beta})=\nonumber\\
&&(A_{\alpha:\rho:\sigma} - A_{\alpha:\sigma:\rho})_{:\beta}+\mathrm{(cyc.~perm.)}
=(R^{\tau}_{\alpha\rho\sigma}A_{\tau})_{:\beta}
+\mathrm{(cyc.~perm.)}=\nonumber\\
&&R^{\tau}_{\alpha\rho\sigma} A_{\tau:\beta} +
R^{\tau}_{\beta\rho\sigma} A_{\alpha:\tau} +\mathrm{(cyc.~perm.)}
\end{eqnarray}
From this relation we find
\begin{equation}\label{Bainc7}
R^{\tau}_{\alpha\rho\sigma:\beta}A_{\tau}+\mathrm{(cyc.~perm.)}=    R^{\tau}_{\beta\rho\sigma} A_{\alpha:\tau} +\mathrm{(cyc.~perm.)}.
\end{equation}
Finally, from (\ref{Bainc7}), taking into account the cyclic relation (\ref{curv9}), we obtain the famous Bianci identity
\begin{equation}\label{Bainc8}
R^{\tau}_{\alpha\rho\sigma:\beta}+R^{\tau}_{\alpha\sigma\beta:\rho}+
R^{\tau}_{\alpha\beta\rho:\sigma}=0.
\end{equation}

 \section{The Ricci curvature tensor}
If we contract a pair of indexes of the Riemann curvature tensor we will  obtain the tensor of the second rank.
Let us determine the rank two Ricci curvature tensor as follows
\begin{equation}\label{Ric1}
    g^{\rho\sigma}R_{\sigma\alpha\beta\rho}=
    R^{\rho}_{\alpha\beta\rho}=R_{\alpha\beta}.
\end{equation}
All other contractions of two indexes of the Riemann curvature tensor give zero or (up to a sign)  the Ricci tensor. In fact, taking into account relations (\ref{curv15}), we have\footnote{Notice that in literature exist different definitions of the Ricci tensor. For example, in \cite{Landau} the Ricci tensor is determined by the contraction $g^{\sigma\beta}R_{\sigma\alpha\beta\rho}$. It is equal to $-R_{\alpha\rho}$.}
\begin{eqnarray}\label{Ric2}
&&g^{\sigma\alpha}R_{\sigma\alpha\beta\rho}=0,
~~g^{\sigma\beta}R_{\sigma\alpha\beta\rho}=-R_{\alpha\rho},~~
g^{\alpha\beta}R_{\sigma\alpha\beta\rho}=R_{\sigma\rho},\nonumber\\
&&g^{\alpha\rho}R_{\sigma\alpha\beta\rho}=-R_{\sigma\beta},~~
g^{\beta\rho}
R_{\sigma\alpha\beta\rho}=0.
\end{eqnarray}
Further, from (\ref{curv15}) we find
\begin{equation}\label{Ric3}
  g^{\rho\sigma}R_{\sigma\alpha\beta\rho}= g^{
  \sigma\rho}R_{\rho\beta\alpha\sigma}.
\end{equation}
Thus, the tensor Ricci is a symmetric tensor
\begin{equation}\label{Ric4}
R_{\alpha\beta}= R_{\beta\alpha}.
\end{equation}
Contracting two indexes of the Ricci tensor, we obtain  the scalar curvature
\begin{equation}\label{Ric4}
 g^{\alpha\beta} R_{\beta\alpha}=R^{\alpha}_{\alpha} =R.
\end{equation}
From (\ref{curv7}) we find that the Ricci curvature tensor is given be the relation
\begin{equation}\label{RIcci}
R_{\alpha\beta}=\Gamma^{\sigma}_{\alpha\sigma,\beta}-
\Gamma^{\sigma}_{\alpha\beta,\sigma}
+\Gamma^{\tau}_{\alpha\sigma}\Gamma^{\sigma}_{\tau\beta}-
\Gamma^{\tau}_{\alpha\beta}\Gamma^{\sigma}_{\tau\sigma}.
\end{equation}
The Bianci identity for the Ricci curvature tensor plays  a fundamental role in the General Theory of Relativity. In order to obtain this identity let us contract two pairs of indexes in the Bianci identity for the Riemann tensor (\ref{Bainc7}). Taking into account that the covariant derivative of the metric tensor $g_{\alpha\beta}$ is equal to zero, from (\ref{Bainc7}) we obtain the following relation
\begin{equation}\label{Ric5}
(g^{\alpha\beta}R^{\tau}_{\alpha\beta\rho})_{:\tau} +(g^{\alpha\beta}R^{\tau}_{\alpha\rho\tau})_{:\beta} +
(g^{\alpha\beta}R^{\tau}_{\alpha\tau\beta})_{:\rho}=0.
\end{equation}
For the different terms in this relation we obviously have
\begin{eqnarray}\label{Ric6}
 g^{\alpha\beta}R^{\tau}_{\alpha\rho\tau}=
 g^{\alpha\beta}R_{\alpha\rho}&=&R^{\beta}_{\rho},~~
 g^{\alpha\beta}R^{\tau}_{\alpha\tau\beta}= g^{\alpha\beta} g^{\tau\sigma}R_{\sigma\alpha\tau\beta}=-g^{\tau\sigma}
 R_{\sigma\tau}=-R,\nonumber\\~~g^{\alpha\beta}
 R^{\tau}_{\alpha\beta\rho}&=&
g^{\alpha\beta} g^{\tau\sigma}R_{\sigma\alpha\beta\rho}=
 g^{\tau\sigma}R_{\sigma\rho}=R^{\tau}_{\rho}.
\end{eqnarray}
From (\ref{Ric5}) and (\ref{Ric6}) we find
\begin{equation}\label{Ric8}
 2 R^{\tau}_{\rho:\tau} -R_{:\rho}=0.
\end{equation}
Finally,  if we multiply  (\ref{Ric8}) by $g^{\rho\mu}$,  the Bianci identity takes the form
\begin{equation}\label{Ric9}
 (R^{\mu\tau} -\frac{1}{2}g^{\mu\tau}R )_{:\tau} =0.
\end{equation}

\section{The Equation of Motion of a Particle in a Riemann Space (Geodesics)}
We will derived here the equation of motion of a particle in a
curved Riemann space. Let $x^{\alpha}(s)$ be a coordinate of a particle on a trajectory. We assume that the
velocity of a particle $u^{\alpha}$ is a timelike vector and  choose a proper time $s$ as a parameter. We have $u^{\alpha}=\frac{dx^{\alpha}}{d s}$. Taking into account that $ds^{2}=dx^{\alpha}dx_{\alpha}$ we have
\begin{equation}\label{Geo1}
    u^{\alpha}u_{\alpha}=1.
\end{equation}
 Let us assume that along the trajectory a velocity is changing by the parallel displacement. We have in this case
\begin{equation}\label{Geo2}
u^{\alpha}(s+ds)=u^{\alpha}(s)+\delta u^{\alpha}(s),
\end{equation}
where
\begin{equation}\label{Geo3}
\delta u^{\alpha}=-\Gamma^{\alpha}_{\beta\rho}u^{\beta} dx^{\rho}=-\Gamma^{\alpha}_{\beta\rho}u^{\beta}u^{\rho}ds.
\end{equation}
From (\ref{Geo2}) and (\ref{Geo3}) we find the following equation \begin{equation}\label{Geo4}
\frac{du^{\alpha}}{ds}+\Gamma^{\alpha}_{\beta\rho}u^{\beta}u^{\rho}=0.
\end{equation}
We will show now that (\ref{Geo4}) is the equation of motion
of a particle in a curved Riemann space.

The  equation of motion of a particle follows from the variational principle
\begin{equation}\label{Geo7}
\delta\int^{Q}_{P}ds=0,
\end{equation}
where $P$ and $Q$ are fixed points on the trajectory.
From
\begin{equation}\label{Geo8}
ds^{2}=g_{\alpha\beta}dx^{\alpha}dx^{\beta}
\end{equation}
we find
\begin{equation}\label{Geo9}
\delta ds^{2}=2ds\delta ds=\delta g_{\alpha\beta}dx^{\alpha}dx^{\beta} +2g_{\alpha\beta}dx^{\alpha}\delta dx^{\beta}.
\end{equation}
Further, using  $\delta dx^{\alpha}=d\delta x^{\alpha}$, we have
\begin{equation}\label{Geo10}
\delta\int^{Q}_{P}ds= \int^{Q}_{P}[\frac{1}{2}g_{\alpha\beta,\rho}u^{\alpha}u^{\beta}\delta x^{\rho} -\frac{d}{ds}(g_{\alpha\rho}u^{\alpha})]\delta x^{\rho}ds +\int^{Q}_{P}d(g_{\alpha\rho}u^{\alpha}\delta x^{\beta})
\end{equation}
Taking into account that $\delta x^{\rho}(P)=\delta x^{\rho}(Q)=0$, from (\ref{Geo7}) and (\ref{Geo10}) we obtain the equation
\begin{eqnarray}\label{Geo11}
\delta\int^{Q}_{P}ds &=&
\int^{Q}_{P}[\frac{1}{2}g_{\alpha\beta,\rho}u^{\alpha}u^{\beta}     -g_{\alpha\rho,\beta}u^{\alpha}u^{\beta}-g_{\alpha\rho}
\frac{du^{\alpha}}{ds}]\delta x^{\rho}ds\nonumber\\ &=&-\int^{Q}_{P}[g_{\alpha\rho}
\frac{du^{\alpha}}{ds}+
\Gamma_{\rho\alpha\beta}u^{\alpha}u^{\beta}]\delta x^{\rho}ds=0,
\end{eqnarray}
where
\begin{equation}\label{Geo12}
    \Gamma_{\rho\alpha\beta}=\frac{1}{2}(g_{\rho\alpha,\beta}+
g_{\rho\beta,\alpha}-g_{\alpha\beta,\rho})
\end{equation}
is  the Christoffel symbol.

Thus, from variational principle we find the equation
\begin{equation}\label{Geo13}
 g_{\alpha\rho}
\frac{du^{\alpha}}{ds}+
\Gamma_{\rho\alpha\beta}u^{\alpha}u^{\beta}=0.
\end{equation}
If we multiply this equation by $g^{\sigma\rho}$ we obviously come to the equation (\ref{Geo4}).

The  equation (\ref{Geo4}) can be written in the form
\begin{equation}\label{Geo14}
\frac{d^{2}x^{\alpha}}{d s^{2}}=-\Gamma^{\alpha}_{\beta\rho}\frac{dx^{\beta}}{ds}
  \frac{dx^{\rho}}{ds}.
\end{equation}
Hence, the motion of the particle in curved Riemann space is determined  by the Christoffel symbols,  $-m\Gamma^{\alpha}_{\beta\rho}u^{\beta}u^{\rho}$ ($m$ is a mass) is a force and components of the metric tensor play a role of potential.

Summarizing, between two fixed points in a curved Riemann space a particle is moving along the shortest trajectory. Such track of a particle is called geodesic. As we will see later, curvature of a space is determined by the gravitational field. Thus, in a gravitational field a particle is moving along the geodesic which is determined by the equation (\ref{Geo4}).

\section{ On the Equivalence Principle}

Since 1907 for Einstein the {\em Equivalence Principle}  was a guide to the Theory of General Relativity. According to the Equivalence Principle  in any infinitesimal region of a four-dimensional space, in which the gravitational force can be considered as a constant, exist such a coordinate system  in which the  gravitational force does not affect movement of a body and all other physical processes. This principle is based on the equality of inertial and gravitational masses which ensure that in the gravitational field all bodies have the same acceleration.

In such a system  and in all other systems, which can be obtained from it by the Lorenz transformation,  all physical laws have the same form. The square of an invariant distance between points $\xi^{\mu}+d\xi^{\mu}$ and $\xi^{\mu}$  in cartesian coordinates is given by
\begin{equation}\label{41}
ds^{2}=\eta_{\mu\nu}d\xi^{\mu}d\xi^{\nu},
\end{equation}
where
\begin{equation}\label{41a}
\eta_{00}=1,~~\eta_{ii}=-1, ~~\eta_{\mu\nu}=0~~(\mu\neq\nu).
\end{equation}
Let us consider  {\em any other coordinate system}. Coordinates of the same point in such a system  $x^{\alpha}$ are functions of $\xi^{\beta}$. For invariant $ds^{2}$  we find
\begin{equation}\label{43}
ds^{2}=\eta_{\mu\nu}\frac{\partial \xi^{\mu}}{\partial x^{\alpha}}\frac{\partial \xi^{\nu}}{\partial x^{\beta}}dx^{\alpha}dx^{\beta}=g_{\alpha\beta}dx^{\alpha}dx^{\beta}.
\end{equation}
Here
\begin{equation}\label{44}
g_{\alpha\beta}=\eta_{\mu\nu}\frac{\partial \xi^{\mu}}{\partial x^{\alpha}}\frac{\partial \xi^{\nu}}{\partial x^{\beta}}
\end{equation}
is the metric tensor.

The equation of the free motion of a particle in the initial inertial Galileo system  (no gravitational force) is given by
\begin{equation}\label{45}
    \frac{d^{2}\xi^{\mu}}{ds^{2}}=0.
\end{equation}
We will shaw now that in any other system  Eq.(\ref{45}) become the equation for geodesic (\ref{Geo4}), considered in the previous section. From (\ref{45}) we have
\begin{equation}\label{46}
\frac{d^{2}\xi^{\mu}}{ds^{2}}=\frac{d}{ds}
(\frac{\partial\xi^{\mu} }{\partial x^{\rho}}\frac{dx^{\rho}}{ds})=
\frac{\partial^{2}\xi^{\mu}}{\partial x^{\rho}\partial x^{\beta} }\frac{dx^{\rho}}{ds}
\frac{dx^{\beta}}{ds}+
\frac{\partial\xi^{\mu}}{\partial x^{\rho}}\frac{d^{2}x^{\rho}}{ds^{2}}=0.
 \end{equation}
Let us multiply this equation by $\frac{\partial x^{\alpha}}{\partial\xi^{\mu}}$. Taking into account that $\frac{\partial\xi^{\mu}}{\partial x^{\rho}}\frac{\partial x^{\alpha}}{\partial\xi^{\mu}}=\frac{\partial x^{\alpha}}{\partial\xi^{\rho}}=
\delta_{\rho}^{\alpha}$, we find
\begin{equation}\label{47}
 \frac{d^{2}x^{\alpha}}{ds^{2}} +
 \frac{d^{2}\xi^{\mu}}{dx^{\rho}dx^{\beta}}
 \frac{dx^{\alpha}}{d\xi^{\mu}}
 \frac{dx^{\rho}}{ds}\frac{dx^{\beta}}{ds}=0.
\end{equation}
Let us introduce the notation
\begin{equation}\label{48}
\frac{d^{2}\xi^{\mu}}{dx^{\rho}dx^{\beta}}
 \frac{\partial x^{\alpha}}{\partial\xi^{\mu}}=\bar{\Gamma}_{\rho\beta}^{\alpha},
\end{equation}
We will show now that $\bar{\Gamma}_{\rho\beta}^{\alpha}$ is the Cristoffel symbol, determined by the relation (\ref{Cristo6}).

In fact, taking into account that
$\eta_{\mu\nu}=\eta_{\nu\mu}$, from  Eq. (\ref{44}) we find
\begin{equation}\label{50}
\frac{\partial g_{\alpha\beta}}{\partial x^{\rho}}=
\eta_{\mu\nu}\frac{\partial^{2} \xi^{\mu}}{\partial x^{\rho}\partial x^{\alpha}}
\frac{\partial \xi^{\nu}}{\partial x^{\beta}}+\eta_{\mu\nu}
\frac{\partial \xi^{\mu}}{\partial x^{\alpha}}\frac{\partial^{2} \xi^{\nu}}{\partial x^{\rho}\partial x^{\beta}}=
 \eta_{\mu\nu}\frac{\partial^{2} \xi^{\mu}}{\partial x^{\rho}\partial x^{\alpha}}
\frac{\partial \xi^{\nu}}{\partial x^{\beta}}+(\alpha\rightleftarrows\beta).
\end{equation}
From (\ref{44}) and (\ref{48}) it follows that
\begin{equation}\label{52}
 \bar{\Gamma}^{\sigma}_{\rho\alpha}g_{\sigma\beta}=
  \frac{\partial^{2} \xi^{\mu}}{\partial x^{\rho}\partial x^{\alpha}}\frac{\partial \xi^{\sigma}}{\partial x^{\mu}} \eta_{\tau\nu}\frac{\partial \xi^{\tau}}{\partial x^{\sigma}}
 \frac{\partial \xi^{\nu}}{\partial x^{\beta}}= \eta_{\mu\nu} \frac{\partial^{2} \xi^{\mu}}{\partial x^{\rho}\partial x^{\alpha}}\frac{\partial \xi^{\nu}}{\partial x^{\beta}}.
\end{equation}
Thus, we find the relation
\begin{equation}\label{53}
\frac{\partial g_{\alpha\beta}}{\partial x^{\rho}}=\bar{\Gamma}^{\sigma}_{\rho\alpha}g_{\sigma\beta}+
\bar{\Gamma}^{\sigma}_{\rho\beta}g_{\sigma\alpha}.
\end{equation}
 If we  change in (\ref{53}) $\alpha\leftrightarrows\rho$ and $\beta\leftrightarrows\rho$ we obtain two additional  relations
\begin{equation}\label{54}
\frac{\partial g_{\rho\beta}}{\partial x^{\alpha}}=    \bar{\Gamma}^{\sigma}_{\alpha\rho}g_{\sigma\beta}+
\bar{\Gamma}^{\sigma}_{\alpha\beta}g_{\sigma\rho}
\end{equation}
and
\begin{equation}\label{55}
\frac{\partial g_{\alpha\rho}}{\partial x^{\beta}}=\bar{\Gamma}^{\sigma}_{\beta\alpha}g_{\sigma\rho}+
\bar{\Gamma}^{\sigma}_{\beta\rho}g_{\sigma\alpha}.
\end{equation}
It is obvious from (\ref{48}) that $\bar{\Gamma}^{\sigma}_{\alpha\rho}=\bar{\Gamma}^{\sigma}_{\rho\alpha}$.  From (\ref{53}), (\ref{54}) and (\ref{55}) we find that $\bar{\Gamma}^{\sigma}_{\alpha\rho}$ is given by the relation
\begin{equation}\label{56}
\bar{\Gamma}^{\sigma}_{\alpha\rho}g_{\sigma\beta}=
\bar{\Gamma}_{\beta\alpha\rho}=
\frac{1}{2}(\frac{\partial g_{\alpha\beta}}{\partial x^{\rho}}+\frac{\partial g_{\rho\beta}}{\partial x^{\alpha}}-\frac{\partial g_{\alpha\rho}}{\partial x^{\beta}}).
\end{equation}
Comparing this relation with (\ref{Cristo6}) we conclude that $\bar{\Gamma}_{\beta\alpha\rho}=\Gamma_{\beta\alpha\rho}$, where $\Gamma_{\beta\alpha\rho}$ is the standard Christoffell index. Thus, from (\ref{47}) we have
\begin{equation}\label{57}
 \frac{d^{2}x^{\alpha}}{ds^{2}} +\Gamma_{\rho\beta}^{\alpha}\frac{dx^{\rho}}{ds}\frac{dx^{\beta}}{ds}=0    \end{equation}
This equation is the equation for geodesics (\ref{Geo4}) which describes the motion of a body in curved Riemann space (gravitational field). We have demonstrated here that non inertial systems are equivalent to a gravitational field which is determined by the metric tensor.

\section{Einstein Equations}
In 1915 after many years of  various attempts Einstein finally formulated equations of the General Relativity. His basic requirement was that equations of the General Theory of Relativity had to be invariant under a general transformation of coordinates in a curved Riemann space. He assumed that the conserved energy-momentum tensor of matter (and radiation) $T^{\alpha\beta}$ was proportional to a tensor $R^{\alpha\beta}+a g^{\alpha\beta}R$ formed by the Ricci tensor.

We have seen before that in a Riemann space the Bianci identity
\begin{equation}\label{Equ1}
 (R^{\alpha\beta} -\frac{1}{2}g^{\alpha\beta}R )_{:\beta} =0.
\end{equation}
must be satisfied (see the section 5.). Thus, we have $a=-\frac{1}{2}$. The Einstein equations took the form
\begin{equation}\label{Equ2}
R^{\alpha\beta}-\frac{1}{2}g^{\alpha\beta}R=-8\pi G T^{\alpha\beta}.
\end{equation}
Here $G$ is the gravitational constant.\footnote{In the system $\hbar=c=1$ $G$ has dimension $M^{-2}$ and the curvature tensor has dimension $M^{2}$. Thus, the tensor $T^{\alpha\beta}$ has dimension $M^{4}$.} The coefficient $-8\pi G $ was chosen in order to ensure the correct  Newtonian approximation.

From (\ref{Equ2}) follows that due to Bianci identity  the equation
\begin{equation}\label{Equ4}
 T^{\alpha\beta}_{:\beta}=0,
\end{equation}
which ensure conservation of energy and momentum, is {\em contained in the Einstein equations}.

In the empty space (no matter and  no fields, except the gravitational field) we have
\begin{equation}\label{Equ5}
R^{\alpha\beta} -\frac{1}{2}g^{\alpha\beta}R  =0.
\end{equation}
Contracting  indexes in (\ref{Equ5}) we find  $R-2R=-R=0$. Thus, the Einstein equation in the empty space can be written in the  form\footnote{This does not mean that such a space is flat. Space is flat only if the Riemann tensor is equal to zero.}
\begin{equation}\label{Equ6}
R^{\alpha\beta} =0.
\end{equation}
In 1917 Einstein included in the equations of the General Relativity the tensor
\begin{equation}\label{Equ7}
\Lambda g^{\alpha\beta},
\end{equation}
where $\Lambda$ is a constant, which is called the cosmological constant. The equations of General Relativity took the form
\begin{equation}\label{Equ5}
R^{\alpha\beta}-\frac{1}{2}g^{\alpha\beta}R +
\Lambda g^{\alpha\beta}=-8\pi G T^{\alpha\beta}.
\end{equation}
 From the condition $g^{\alpha\beta}_{:\beta}=0$ (see (\ref{covar7})) follows that the equation (\ref{Equ4}) is contained in the equations of the General Relativity also in the case of the cosmological constant.

 From the Einstein equations (\ref{Equ5}) follows that evolution of the Universe is determined not only by attractive gravitational force but also by a proportional to $\Lambda$ repulsive force. From analysis of existing cosmological data follows that about 70\% of the density of the Universe at present time is due to the cosmological constant (or dark energy)

 The Einstein equations are equations for the metric tensor $g^{\alpha\beta}$ which play a role of a potential of  the gravitational field. From  (\ref{RIcci}) follows that Einstein equations are the second order equations for the potential. Notice that ten components of  $g_{\alpha\beta}$ describe not only potential but also coordinate system.

\section{Energy-Momentum Tensor}
Here we will consider an energy-momentum tensor for continuously distributed  matter. The velocity of  an infinitesimal element of the matter with coordinates $x^{\alpha}$ is given by
\begin{equation}\label{energymom1}
 u^{\alpha}=\frac{dx^{\alpha}}{ds}.
\end{equation}
Taking into account that
\begin{equation}\label{energymom2}
ds^{2}=g_{\alpha\beta}dx^{\alpha}dx^{\beta}
\end{equation}
we have
\begin{equation}\label{energymom3}
g_{\alpha\beta}u^{\alpha}u^{\beta}=1.
\end{equation}
In order to built the conserved energy-momentum tensor we will use the relation
\begin{equation}\label{energymom4}
\sqrt{-g}_{,\beta}=\sqrt{-g}~\Gamma^{\alpha}_{\beta\alpha}.
\end{equation}
Here $g$ is the determinant of the metric tensor $g_{\alpha\beta}$ ($g<0$).

 Using (\ref{energymom4}) for any vector $A^{\alpha}$ we find the following relation
\begin{equation}\label{energymom5}
 \sqrt{-g} A^{\alpha}_{:\alpha}=\sqrt{-g}A^{\alpha}_{,\alpha}
 +\sqrt{-g}\Gamma^{\alpha}_{\sigma\alpha}A^{\sigma}=
 \sqrt{-g}A^{\alpha}_{,\alpha}+\sqrt{-g}_{,\alpha}A^{\alpha}
 =(\sqrt{-g}~A^{\alpha})_{,\alpha}.
\end{equation}
Let us introduce now the scalar quantity $\rho(x)$ and assume that the vector $\sqrt{-g}~\rho u^{\alpha}$ is conserved
\begin{equation}\label{energymom6}
(\sqrt{-g}~\rho u^{\alpha})_{,\alpha}=0.
\end{equation}
 From this relation follows that $\sqrt{-g}~\rho u^{0}$ is the density of matter and $\sqrt{-g}~\rho u^{i}$ is the flux of matter.

We will choose the symmetric energy-momentum tensor in the form
\begin{equation}\label{energymom7}
T^{\alpha\beta}=\rho u^{\alpha}u^{\beta}.
\end{equation}
Let us calculate the covariant derivative of this tensor. We have
\begin{equation}\label{energymom9}
T^{\alpha\beta}_{:\beta}=u^{\alpha} (\rho u^{\beta})_{:\beta} +
u^{\alpha}_{:\beta}\rho u^{\beta}.
\end{equation}
From (\ref{energymom5}) and (\ref{energymom6}) we find
\begin{equation}\label{energymom10}
\sqrt{-g}(\rho u^{\beta})_{:\beta}=(\sqrt{-g}\rho u^{\beta})_{,\beta}=0.
\end{equation}
Further, taking into account that $\frac{du^{\alpha}}{ds}=\frac{du^{\alpha}}{dx^{\beta}}u^{\beta}$, from the equation for the geodesic (\ref{Geo4}) we have
\begin{equation}\label{energymom11}
    u^{\alpha}_{:\beta} u^{\beta}=0.
\end{equation}
From (\ref{energymom10}) and (\ref{energymom11}) follows that the tensor $T^{\alpha\beta}$, given by the expression (\ref{energymom7}), is conserved
\begin{equation}\label{energymom13}
T^{\alpha\beta}_{:\beta}=0.
\end{equation}

\section{Newtonian approximation}
As we have seen in the section 7, in the gravitational field  particles are moving along the geodesic. The equation of motion has the form
\begin{equation}\label{Na1}
\frac{du^{\alpha}}{ds} =-\Gamma^{\alpha}_{\beta\rho}u^{\beta}u^{\rho}.
\end{equation}
In this section we consider the movement of a non relativistic particle in a  weak, stationary gravitational field. For the stationary field the metric tensor  does not depend on time
\begin{equation}\label{Na2}
 g_{\alpha\beta},_{0} \simeq 0
\end{equation}
and
\begin{equation}\label{Na3}
 g_{i 0} \simeq 0.
\end{equation}
Taking into account that $g^{\alpha\beta}g_{\beta\rho}=\delta^{\alpha}_{\rho}$ we also have
\begin{equation}\label{Na4}
g^{i 0} \simeq 0, \quad g^{0 0}g_{0 0}\simeq1,\quad g^{ik}g_{kl}\simeq\delta^{i}_{l}.
\end{equation}
In the liner over velocity approximation from (\ref{Na3}) we find
\begin{equation}\label{Na5}
\frac{du^{i}}{ds}
=-g^{ik}\Gamma_{k\beta\rho}u^{\beta}u^{\rho}
\simeq-g^{ik}\Gamma_{k 0 0}u^{0}u^{0}.
\end{equation}
From (\ref{Cristo6}) and  (\ref{Na2}) we easily find that
\begin{equation}\label{Na6}
 \Gamma_{k 0 0}\simeq-\frac{1}{2}g_{00,k}.
\end{equation}
Further, in the linear approximation we have
\begin{equation}\label{Na7}
 \frac{du^{i}}{ds}=\frac{du^{i}}{dx^{\alpha}}u^{\alpha}\simeq \frac{du^{i}}{dx^{0}}u^{0}
\end{equation}
and
\begin{equation}\label{Na8}
g_{\alpha\beta}u^{\alpha}u^{\beta}\simeq g_{00}(u^{0})^{2}= 1,\quad u^{0}\simeq g^{-1/2}_{00}.
\end{equation}
From (\ref{Na5})-(\ref{Na8}) we obtain the following equation
of the motion of non relativistic particle in a weak gravitational field
\begin{equation}\label{Na8a}
 \frac{du^{i}}{dx^{0}}= g^{ik}(g^{1/2}_{00})_{,k}.
\end{equation}
From this equation we conclude that $\sqrt{g_{00}}$ plays a role of a potential of a particle in the gravitational field. In fact, the Einstein equation in the gravitational field has a form \begin{equation}\label{Na9}
    R_{\alpha\beta}=0.
\end{equation}
We are considering  a weak gravitational field. For such a field
the metric tensor is approximately constant and Cristoffel symbols are small. Neglecting quadratic over Cristoffel symbols terms in the Ricci curvature tensor we have
\begin{equation}\label{Na10}
\Gamma^{\rho}_{\alpha\rho,\beta}-
\Gamma^{\rho}_{\alpha\beta,\rho}\simeq 0.
\end{equation}
Using (\ref{Cristo6}), from this equation we find
\begin{equation}\label{Na11}
  g^{\rho\tau}  [g_{\tau\rho,\alpha\beta}-g_{\alpha\rho,\tau\beta}
-g_{\tau\beta,\alpha\rho}+g_{\alpha\beta,\tau\rho}]=0.
\end{equation}
 Let us put in this equation $\alpha=\beta=0$. Taking into account (\ref{Na2}), we find that $g_{00}$ satisfies  the Laplace equation
\begin{equation}\label{Na12}
 g^{ik} g_{00,ik}=0.
\end{equation}
If there is no gravitational field we have $g_{00}=1$. In the case of the weak gravitational field we can present $g_{00}$ in the form
\begin{equation}\label{Na14}
    g_{00}=1+2V,
\end{equation}
where $V$ is a small quantity which satisfies the Laplace equation
\begin{equation}\label{Na15}
\nabla^{2}V=0.
\end{equation}
From (\ref{Na8a}) and (\ref{Na14}) we find the equation
\begin{equation}\label{Na16}
\frac{d\vec{u}}{dx^{0}}=-\nabla V
\end{equation}
Comparing this equation with the equation of motion of a particle in the gravitational filed, we conclude
that $V$ can be identified with the Newton potential
\begin{equation}\label{Na17}
    V=-G\frac{M}{r},
\end{equation}
where G is the gravitational constant and $M$ is a mass at the origin.
\section{Spherically Symmetrical Gravitational Field. The Schwarzschild Solution.}
In this section we will consider {\em the static, spherically symmetrical gravitational field} produced by a spherically symmetrical
body at rest.

For the static, spherically symmetrical field we have
\begin{equation}\label{Schw1}
 g_{\alpha\beta,0}=0,\quad g_{0 i}=0.
\end{equation}
Let us choose the spherical  coordinates $r$, $\theta$, $\phi$. The most general expression for $ds^{2}$ in  spherically symmetrical static case has the form
\begin{equation}\label{Schwarz2}
ds^{2}=U(r) dt^{2}-V(r)dr^{2}-W(r)
 (d\theta^{2}+\sin^{2}\theta d\phi^{2}),
\end{equation}
where $U$, $V$ and $W$ are functions of $r$.  The coordinate $r$ can be arbitrary chosen. We can determine  $r$ in such a way that $W(r)= r^{2}$.\footnote{In this case the length of the circle with the center at the  center of coordinates is equal to $2\pi r$.} It is convenient to present functions $U$ and $V$ in the form $U(r)=e^{2\nu(r)}$ and $V(r)=e^{2\lambda(r)}$. The interval $ds^{2}$ takes the form
\begin{equation}\label{Schwarz3}
 ds^{2}=e^{2\nu}dt^{2}-e^{2\lambda}dr^{2}-r^{2}(d\theta^{2}+\sin^{2}\theta d\phi^{2}),
\end{equation}
We have $x^{0}=t$, $x^{1}=r$, $x^{2}=\theta$, $x^{3}=\phi$.
 From (\ref{Schwarz3}) follows that nonzero components of the metric tensor $g_{\alpha\beta}$ are given by
\begin{equation}\label{Schwarz5}
g_{00}=e^{2\nu},~~g_{11}=-e^{2\lambda},~~g_{22}
=-r^{2},~~g_{33}=-r^{2}\sin^{2}\theta.
\end{equation}
Taking into account that $g^{\alpha\beta}g_{\beta\rho}=\delta^{\alpha}_{\rho}$ we find
\begin{equation}\label{Schwarz6}
g^{00}=e^{-2\nu},~~g^{11}=-e^{-2\lambda},~~g^{22}=-r^{-2},~~ g^{33}=-r^{-2}\sin^{-2}\theta.
\end{equation}
Let us now calculate  the  Christoffel symbols $\Gamma^{\alpha}_{\beta\gamma}$.  From (\ref{Cristo6}), (\ref{Schwarz5}) and (\ref{Schwarz6}) we find
\begin{eqnarray}\label{Schwarz7}
\Gamma^{0}_{10}=\nu',~~\Gamma^{1}_{00}&=&       e^{2(\nu-\lambda)}\nu',~~\Gamma^{1}_{11}=\lambda',~~\Gamma^{2}_{12}=\Gamma^{3}_{13}=r^{-1},~~
\Gamma^{1}_{22}=-re^{-2\lambda},\nonumber\\
\Gamma^{3}_{23}&=&\frac{\cos\theta}{\sin\theta},~~\Gamma^{1}_{33}=-r\sin^{2}\theta e^{-2\lambda},~~\Gamma^{2}_{33}=-\sin\theta\cos\theta.
\end{eqnarray}
Other components of the the  Christoffel symbols are equal to zero.

From (\ref{curv7}), (\ref{Ric1}) and (\ref{Schwarz7}) for nonzero components of the Ricci curvature tensor we find the following expressions
\begin{equation}\label{Schwarz8}
  R_{00}=(\lambda'\nu'-\frac{2\nu'}{r}-\nu''-\nu'^{2})
  ~e^{2(\nu-\lambda)},
\end{equation}
\begin{equation}\label{Schwarz9}
R_{11}=\nu''-\frac{2\lambda'}{r}-\lambda'\nu'+\nu'^{2},
\end{equation}
\begin{equation}\label{Schwarz10}
R_{22}=(+1+r\nu'-r\lambda')~e^{-2\lambda} -1
\end{equation}
and
\begin{equation}\label{Schwarz11}
R_{33} =R_{22}\sin^{2}\theta.
\end{equation}
The Einstein equations for the gravitational field outside of a body which produce the field  have a form
\begin{equation}\label{Schwarz12}
R_{\alpha\beta}=0.
\end{equation}
From (\ref{Schwarz8}), (\ref{Schwarz9}) and (\ref{Schwarz12}) we obtain the following equation
\begin{equation}\label{Schwarz13}
    \nu'+\lambda'=0.
\end{equation}
From this equation we find
\begin{equation}\label{Schwarz14}
\nu(r)+\lambda(r)=C,
\end{equation}
where $C$ is a constant. At $r\to \infty$ the space is flat and $\nu=\lambda=0$. Thus, we have $C=0$ and
\begin{equation}\label{Schwarz15}
\lambda(r)=-\nu(r).
\end{equation}
From  (\ref{Schwarz10}) and (\ref{Schwarz12}) we have
\begin{equation}\label{Schwarz16}
(1+2r\nu')~e^{2\nu}=(re^{2\nu})'=1.
\end{equation}
Thus, we find
\begin{equation}\label{Schwarz17}
e^{2\nu}=1+ \frac{C_{1}}{r},
\end{equation}
where $C_{1}$ is a constant. At large $r$ the Newton approximation is valid and $g_{00}$ is given by
\begin{equation}\label{Schwarz18}
 g_{00}=1+2V,\quad V=-G\frac{M}{r},
\end{equation}
where $M$ is the mass of the body which produce the gravitational field. Comparing (\ref{Schwarz17}) and (\ref{Schwarz18}) we conclude that $C_{1}=-2GM$ and
\begin{equation}\label{Schwarz20}
g_{00}=e^{2\nu}=1-G\frac{2M}{r}.
\end{equation}
 We also have $g_{11}=e^{-2\nu}=(1-G\frac{2M}{r})^{-1}$. Thus for the static, spherically symmetrical gravitational field we find the following Schwarzschield solution of the Einstein equations
\begin{equation}\label{Schwarz21}
ds^{2}=(1-G\frac{2M}{r}) dt^{2}-(1-G\frac{2M}{r})^{-1}dr^{2}-r^{2}(d\theta^{2}+\sin^{2}\theta d\phi^{2}).
\end{equation}
The Schwarzschield metric modify the Newton theory of the motion of planets around the sun. It perfectly explains  precession of the perihelion of Mercury.

Let us notice that from (\ref{Schwarz21}) it follows that the distance between points $r_{2}$ and $r_{1}$ at the same radius is larger than $(r_{2}-r_{1})$. In fact, we have
\begin{equation}\label{Schwarz22}
    \int^{r_{2}}_{r_{1}}\frac{dr}{(1-G\frac{2M}{r})}>(r_{2}-r_{1}).
\end{equation}
We also have $g_{00}\leq 1$. For the proper time we find
\begin{equation}\label{Schwarz23}
d\tau =\sqrt{g_{00}}~dt \leq dt
\end{equation}
From (\ref{Schwarz20}) and (\ref{Schwarz23}) we conclude that at infinity $t$ and $\tau$ are the same but at a finite distance there is a slowdown of time with respect to a time at infinity.

\section{Gravitational Waves}
In this section we will discuss gravitational waves, one of the very important recently observed consequence of the General Relativity.
For the weak gravitational field in the empty space the Einstein equations (in the linear approximation) have the form
 \begin{equation}\label{GW1}
\Gamma^{\rho}_{\alpha\rho,\beta}-
\Gamma^{\rho}_{\alpha\beta,\rho}\simeq 0.
\end{equation}
Using (\ref{Cristo6}), from this equation we have
\begin{equation}\label{GW2}
  g^{\tau\rho}  [g_{\tau\rho,\alpha\beta}-g_{\tau\alpha,\rho\beta}
-g_{\tau\beta,\alpha\rho}+g_{\alpha\beta,\tau\rho}]=0.
\end{equation}
It is convenient to use harmonic coordinates which provide the closest approximation to the flat case. In these coordinates the following condition holds
\begin{equation}\label{GW3}
 g^{\tau\rho} \Gamma^{\alpha}_{\tau\rho}=0
\end{equation}
From (\ref{GW3}) we find
\begin{equation}\label{GW4}
g^{\tau\rho}(g_{\alpha\tau,\rho}-\frac{1}{2} g_{\tau\rho,\alpha})=0.
\end{equation}
Let us differentiate this equation over $x^{\beta}$. In the linear approximation we have
\begin{equation}\label{GW4}
g^{\tau\rho}(g_{\tau\alpha,\rho\beta}-\frac{1}{2} g_{\tau\rho,\alpha\beta})=0.
\end{equation}
 If we perform in  (\ref{GW4}) the change $\alpha\leftrightarrows\beta$,  we find
\begin{equation}\label{GW5}
g^{\tau\rho}(g_{\tau\beta,\rho\alpha}-\frac{1}{2} g_{\tau\rho,\alpha\beta})=0.
\end{equation}
Adding (\ref{GW2}), (\ref{GW4}) and (\ref{GW5}) we obtain the d'Alambert equation for each $g_{\alpha\beta}$:
\begin{equation}\label{GW6}
 g^{\tau\rho}g_{\alpha\beta,\tau\rho} =0.
\end{equation}
The solution of the equation (\ref{GW6}) are gravitational waves traveling with the speed of light.

  Let us consider gravitational waves  moving in one direction, determined by a vector $l_{\alpha}$, which satisfies the equation $g^{\alpha\beta}l_{\alpha}l_{\beta}=0$. In this case the metric tensor depends on $lx=l_{\alpha}x^{\alpha}$. We have
\begin{equation}\label{GW7}
 g_{\tau\rho,\alpha}=u_{\tau\rho}l_{\alpha},
\end{equation}
where $u_{\tau\rho}=\frac{dg_{\tau\rho}}{dlx}$. For harmonic coordinates from (\ref{GW4}) we find
\begin{equation}\label{GW8}
g^{\tau\rho}u_{\tau\alpha}l_{\rho}=
\frac{1}{2}g^{\tau\rho}u_{\tau\rho}l_{\alpha}=
\frac{1}{2}ul_{\alpha},
\end{equation}
where $u=g^{\tau\rho}u_{\tau\rho}$. This equation can be rewritten in the form
\begin{equation}\label{GW9}
    u^{\rho}_{\alpha}l_{\rho}=\frac{1}{2}ul_{\alpha}
\end{equation}
or
\begin{equation}\label{GW10}
(u^{\tau\rho}-\frac{1}{2}g^{\tau\rho}u)l_{\rho}=0.
\end{equation}
Further from (\ref{Cristo6}) and (\ref{GW7}) we find
\begin{equation}\label{GW11}
    \Gamma^{\alpha}_{\tau\beta}=\frac{1}{2}(u^{\alpha}_{\tau}l_{\beta}
+u^{\alpha}_{\beta}l_{\tau}-u_{\tau\beta}l^{\alpha}).
\end{equation}
  We will consider now the energy-momentum tensor  of the gravitational waves. The invariant action is given by
\begin{equation}\label{GW12}
    I=\int \mathcal{L}\sqrt{-g}d^{4}x,
\end{equation}
where the action density $\mathcal{L}$ is equal to
\begin{equation}\label{GW13}
    \mathcal{L}=g^{\tau\rho}(\Gamma^{\beta}_{\tau\rho}
    \Gamma^{\alpha}_{\beta\alpha}-\Gamma^{\alpha}_{\tau\beta}
    \Gamma^{\beta}_{\rho\alpha}).
\end{equation}
In the harmonic coordinates we have
\begin{equation}\label{GW13}
 \mathcal{L}=-g^{\tau\rho}\Gamma^{\alpha}_{\tau\beta}
    \Gamma^{\beta}_{\rho\alpha}=-\frac{1}{4}(u^{\alpha}_{\tau}l_{\beta}
+u^{\alpha}_{\beta}l_{\tau}-u_{\tau\beta}l^{\alpha})
(u^{\beta}_{\rho}l_{\alpha}
+u^{\beta}_{\alpha}l_{\rho}-u_{\rho\alpha}l^{\beta}).
\end{equation}
Taking into account that    $g^{\alpha\beta}l_{\alpha}l_{\beta}=0$ we conclude that for the gravitational waves moving in one direction $\mathcal{L}=0$.

The (pseudo) energy-momentum tensor $t_{\alpha}^{\beta}$ is determined by the relation
\begin{equation}\label{GW14}
16\pi t_{\tau}^{\rho}\sqrt{-g}=(\Gamma^{\rho}_{\alpha\beta}-
g^{\rho}_{\beta}\Gamma^{\sigma}_{\alpha\sigma})
(g^{\alpha\beta}\sqrt{-g})_{,\tau}-g^{\rho}_{\tau}\mathcal{L}.
\end{equation}
We have
\begin{equation}\label{GW15}
g^{\alpha\beta}_{,\tau}=-g^{\alpha\rho}g^{\beta\sigma}
g_{\rho\sigma,\tau}=
u^{\alpha\beta}l_{\tau}
\end{equation}
and
\begin{equation}\label{GW16}
    \sqrt{-g}_{,\tau}=\frac{1}{2}\sqrt{-g}g^{\alpha\beta}
    g_{\alpha\beta,\tau}=\frac{1}{2}\sqrt{-g}ul_{\tau}
\end{equation}
Thus, we find
\begin{equation}\label{GW17}
(g^{\alpha\beta}\sqrt{-g})_{,\tau}=-
(u^{\alpha\beta}-\frac{1}{2}g^{\alpha\beta}u)\sqrt{-g}l_{\tau}
\end{equation}
From (\ref{GW16}) and (\ref{GW17}) we have
\begin{equation}\label{GW18}
    \Gamma^{\sigma}_{\alpha\sigma}(g^{\alpha\beta}\sqrt{-g})_{,\tau}
=\sqrt{-g}_{,\alpha}(-u^{\alpha\beta}+
\frac{1}{2}g^{\alpha\beta}u)l_{\tau}=\frac{1}{2}\sqrt{-g}ul_{\alpha}
(-u^{\alpha\beta}+
\frac{1}{2}g^{\alpha\beta}u)l_{\tau}.
\end{equation}
Taking into account (\ref{GW10}) we conclude that
\begin{equation}\label{GW19}
\Gamma^{\sigma}_{\alpha\sigma}(g^{\alpha\beta}\sqrt{-g})_{,\tau}=0.    \end{equation}
We have
\begin{equation}\label{GW20}
  16\pi t_{\tau}^{\rho}\sqrt{-g}=
 -\Gamma_{\alpha\beta}^{\rho}(u^{\alpha\beta}-
 \frac{1}{2}g^{\alpha\beta}u)l_{\tau}=-
 \frac{1}{2}(u^{\rho}_{\alpha}l_{\beta}
+u^{\rho}_{\beta}l_{\alpha}-u_{\alpha\beta}l^{\rho})(u^{\alpha\beta}-
 \frac{1}{2}g^{\alpha\beta}u)l_{\tau}
\end{equation}
From (\ref{GW10}) it is easy to see that only the term
\begin{equation}\label{GW21}
 16\pi t_{\tau}^{\rho}\sqrt{-g}=\frac{1}{2}(u_{\alpha\beta}u^{\alpha\beta}-
 \frac{1}{2}u^{2})l^{\rho}l_{\tau}.
\end{equation}
is different from zero and transforms as  tensor. Thus
 in the coordinate system, which is moving in the direction of the vector $l_{\alpha}$, $t_{\tau}^{\rho}\sqrt{-g}$ is the energy-momentum tensor.

Discovery of predicted by the GTR
 gravitational wave signals by LIGO and later Virgo collaborations, awarded by the Nobel Prize in 2017,  is a great success of this theory.

\section{Friedman Equations}
The modern cosmology is based on {\em the Cosmological Principle} which states that at large scales the distribution of Galaxies in the Universe is isotropic and homogeneous. The Cosmological Principle was formulated  by  Einstein in 1917. From the requirement of the isotropy and homogeneity it follows that the metrics of the Universe is the Friedman-Robertson-Walker metrics:
\begin{equation}\label{Cos1}
ds^{2}=dt^{2}-a^{2}(t)(\frac{dr^{2}}{1-kr^{2}}
+r^{2}~(d\theta^{2}+\sin^{2}\theta d\phi^{2})).
\end{equation}
Here $t$ is a cosmological universal time,  $r$ is the comoving distance, $a(t)$ is the scale factor. The distance between any two Galaxies  is given by the relation
\begin{equation}\label{Cos2}
    d(t)=a(t)~r.
\end{equation}
From (\ref{Cos2}) follows
\begin{equation}\label{Cos3}
v(t)=\dot{a}(t)~r,
\end{equation}
where $v(t)=\dot{a}(t)$ is the relative velocity of a Galaxy. From (\ref{Cos2}) and  (\ref{Cos3}) we find the famous Hubble law
\begin{equation}\label{Cos4}
    v(t)=H(t)~d(t),
\end{equation}
where
\begin{equation}\label{Cos5}
H(t)=\frac{\dot{a}(t)}{a(t)}
\end{equation}
is the Hubble parameter.

The parameter $k$ in (\ref{Cos1}) is the  curvature of the space. For the isotropic and homogeneous Universe  $k$ takes three values:
$k=1$ (closed Universe), $k=0$ (flat Universe), ($k=-1$ open Universe).

Comparing (\ref{Cos1}) with the general expression for the interval $ds^{2}=g_{\alpha\beta}dx^{\alpha}dx^{\beta}$ we conclude that
 $$x^{0}=t,~~ x^{1}=r,~~ x^{2}=\theta,~~ x^{3}=\phi$$
 and  nonzero components of the metric tensor are equal to
\begin{equation}\label{Cos6}
g_{00}=1,~~g_{11}=-a^{2}\frac{1}{1-kr^{2}},~~g_{22}=-a^{2}r^{2},~~    g_{33}=-a^{2}r^{2}\sin^{2}\theta.
\end{equation}
 Taking into account relation $g^{\alpha\beta}g_{\beta\rho}=\delta^{\alpha}_{\rho}$ we find
that $g^{00}=1$ and $g^{ii}=g^{-1}_{ii}$.

In order to find components of the Ricci tensor
\begin{equation}\label{1Cos7}
 R_{\alpha\beta}=R^{\rho}_{\alpha\beta\rho}=-
\Gamma^{\rho}_{\alpha\beta,\rho}+\Gamma^{\rho}_{\alpha\rho,\beta}
-\Gamma^{\tau}_{\alpha\beta}\Gamma^{\rho}_{\tau\rho}
+\Gamma^{\tau}_{\alpha\rho}\Gamma^{\rho}_{\tau\beta}
\end{equation}
we need  to calculate the Christoffel symbols
\begin{equation}\label{Cos7}
\Gamma_{\sigma\beta\gamma}=\frac{1}{2}(g_{\sigma\beta,\gamma}
+g_{\sigma\gamma,\beta}-g_{\beta\gamma,\sigma}).
\end{equation}
From (\ref{Cos6}) and (\ref{Cos7}) we find
\begin{eqnarray}\label{Cos8}
\Gamma^{0}_{11}&=&\frac{a\dot{a}}{1-kr^{2}},~~
\Gamma^{0}_{22}=a\dot{a}r^{2},~~
\Gamma^{0}_{33}=a\dot{a}r^{2}\sin^{2}\theta,\nonumber\\
\Gamma^{1}_{01}&=& \Gamma^{1}_{10}=  \Gamma^{2}_{02}= \Gamma^{2}_{20}= \Gamma^{3}_{03} =\Gamma^{3}_{30}=\frac{\dot{a}}{a},\nonumber\\
\Gamma^{1}_{11}&=&\frac{rk^{2}}{1-kr^{2}},~~ \Gamma^{1}_{22}=-r(1-kr^{2}),~~
 \Gamma^{1}_{33}=-(1-kr^{2})r\sin^{2}\theta,\nonumber\\
\Gamma^{2}_{12}&=& \Gamma^{2}_{21}= \Gamma^{3}_{13}= \Gamma^{3}_{31}=\frac{1}{r},\nonumber\\
\Gamma^{2}_{33}&=&-\sin\theta\cos\theta,~~
  \Gamma^{3}_{23}=\Gamma^{3}_{32}=\frac{1}{\tan\theta}.
\end{eqnarray}
For the components of the Ricci tensor from (\ref{1Cos7}) and (\ref{Cos8}) we have
\begin{equation}\label{Cos9}
    R_{00}=3\frac{\ddot{a}}{a},\quad  R_{ik}=\left[\frac{\ddot{a}}{a}+
    2\frac{\dot{a}^{2}}{a^{2}}+2\frac{k}{a^{2}}\right]g_{ik}.
\end{equation}
The Ricci curvature is equal to
\begin{equation}\label{Cas10}
    R=6\left[\frac{\ddot{a}}{a}+
    \frac{\dot{a}^{2}}{a^{2}}+\frac{k}{a^{2}}\right].
\end{equation}
Let turn now to the Einstein equation (\ref{Equ5}). Cosmology is based on the assumption that matter can be considered as an perfect fluid and the energy-momentum tensor of the matter is given by the expression
\begin{equation}\label{Cos11}
 T_{\alpha\beta}=(\rho +p)u_{\alpha}u^{\beta}- p~g_{\alpha\beta}.    \end{equation}
Here $\rho$ and $p$ are the density and pressure of matter
and $u^{\alpha}=\frac{dx^{\alpha}}{ds}$ is the velocity. From the Cosmological Principle (isotropy) it follows that $u=(1,0,0,0)$.
Thus, for $00$ component we find
\begin{equation}\label{Cos12}
 3\frac{\ddot{a}}{a}-3 \left[\frac{\ddot{a}}{a}+
    \frac{\dot{a}^{2}}{a^{2}}+\frac{k}{a^{2}}\right]+\Lambda=-8\pi G\rho
\end{equation}
From (\ref{Cos12}) we obtain the first Friedman equation
\begin{equation}\label{Cos13}
 \frac{\dot{a}^{2}}{a^{2}}=\frac{8\pi }{3}G
 \rho +\frac{1}{3}\Lambda -\frac{k}{a^{2}}.
\end{equation}
Let us consider now $(ii)$ component of the Einstein equation. We have
\begin{equation}\label{Cos14}
\left[\frac{\ddot{a}}{a}+
    2\frac{\dot{a}^{2}}{a^{2}}+2\frac{k}{a^{2}}\right]g_{ii}
-3\left[\frac{\ddot{a}}{a}+
    \frac{\dot{a}^{2}}{a^{2}}+\frac{k}{a^{2}}\right]g_{ii}
+\Lambda g_{ii}=8\pi G pg_{ii}.
\end{equation}
From (\ref{Cos13}) and (\ref{Cos14}) we find the second Freedman equation
\begin{equation}\label{Cos15}
\frac{\ddot{a}}{a}=-\frac{4\pi}{3}G(\rho+3p)+\frac{1}{3}\Lambda.    \end{equation}
Notice that terms in the right-hand side of (\ref{Cos15}) have {\em different signs}: the first gravitational term is negative (attraction) and the second $\Lambda$-term is positive (repulsion). If the second term dominates, the expansion rate of the Universe will be accelerating. As it is well known, we are living now in the accelerating Universe.

\appendix

 \section{ Brief History of the Discovery of the Einstein Equations}

 Here I will discuss a brief history of the A.Einstein equation (for details see, for example, \cite{Pais}). Einstein formulated the Special Relativity in its final form in 1905 in the very first paper on the subject. He wrote more than  ten different papers on the General Relativity before he came to the ultimate formulation of the Equations of the General Relativity in 1915. Einstein moved forward by "trial and error method" during eight years of very intensive work.\footnote{In the letter to Ehrenfest in 1913, apologizing for a long silence, Einstein wrote  "My excuse lies in the literally superhuman efforts which I have devoted to the gravitational problem" (Pais p.223)}

In 1907 Einstein came to an idea of a free-falling cabin which he called "the happiest thought of my life".\footnote{Einstein remembered : "I was sitting in a chair in the patent office at Bern when all of a sudden a thought occurred to me: If a person falls freely he will not feel his own weight. I was startled. This simple thought made a deep impression on me. It impelled me to the theory of gravitation" (Pais, p.179).} For an observer in a free-falling cabin there is no gravitational field (more exactly, because of the equality of inertial and gravitational masses, in the free-falling cabin the gravitational force and  the force of inertia cancel each other.) When several observers are falling in the gravitational field they have no possibility to decide that the gravitational field exist. Einstein thought that this is a powerful argument in favor of necessity of a generalization of the postulate of relativity on non inertial coordinate systems.

The road to the General Theory of Relativity started in 1907 in a review article \cite{Rev} which Einstein wrote by the invitation of the editor of the Jahrbuch. The first part of the review was devoted to the Special Theory of Relativity. In the second part of the review Einstein tried to generalize the principle of relativity  on non inertial systems. He considered uniform gravitational field. In the Newton theory an inertial system in which there is such a field   is equivalent (in terms of  mechanical motion) to the non inertial system with a constant acceleration (and without gravitational field). Einstein postulated that all processes flow in both systems in the same way. He called this postulate {\em the equivalence principle}. On the basis of the equivalence principle Einstein calculated in \cite{Rev} such effects of the gravitational field as the red-shift of spectral lines and bending of the light.

In 1908 Einstein started his academic career. He got a teaching position at the Bern University. In 1909  he  received the position of associate professor of theoretical physics at the University of Z\"urich. In 1911 Einstein  took the position of full professor in the German University at Prague. From  1907 till  1911 he published many papers mainly on quantum theory of light and even took part in some experiment. In Prague he returned to the  gravitation.

Einstein realized that deflection of light in the gravitational field leads to the detectable effect of displacement of the position of stars, observed near the edge of the solar disc.  From the equivalence principle he came to fundamental conclusion that the light velocity depend on the gravitational potential. In 1912 Einstein returned back to Z\"urich. He took the professor position at  ETH (Swiss Federal Institute of Technology).

Toward the end of his Prague time Einstein suggested that the Riemann geometry is a correct mathematical tool for General Theory of Relativity. (``Euclidean geometry must be abandoned if non inertial frames are admitted on the equal footing"). In Zurich Einstein started very fruitful  collaboration with mathematician Marcel Grossmann, his friend from the student years .
In the paper "Outline of a Generalized Theory of Relativity and of a Theory of Gravitation" \cite{EG} Einstein and Grossmann made a significant progress in the  development of a Theory of General Relativity. For the first time they used the Riemann geometry as a mathematical apparatus of a Theory of Gravitation with
the basic idea that gravitational field is determined by the metric tensor $g_{\alpha\beta}$. They  suggested that $g_{\alpha\beta}$ is determined by a energy-momentum tensor and the equation of the General Relativity has the following tensor form
\begin{equation}\label{5}
    X_{\alpha\beta}=k T_{\alpha\beta},
\end{equation}
where $T_{\alpha\beta}$ is a conserved energy-momentum tensor and $X_{\alpha\beta}$ is a second rank tensor determined by quantities which are generated by the metric tensor.

Einstein and Grossmann naturally looked for such $X_{\alpha\beta}$ which in the Newtonian limit of a weak gravitational field gave the Poisson equation
\begin{equation}\label{6}
  \triangle  \phi=4 \pi G \rho.
\end{equation}
The covariant derivative of the metric tensor is equal to zero
\begin{equation}\label{7}
g_{\alpha\beta:\sigma}=0.
\end{equation}
On the basis of this equation it was concluded in \cite{EG}
 ``that the sought for equations will be covariant only with respect to a certain group of transformations..."

Einstein and Grossmann considered the  Ricci curvature tensor $R_{\alpha\beta}$ as a possible candidate for $X_{\alpha\beta}$. However,  they did not managed in \cite{EG} to get Newtonian limit from the Ricci tensor.

Einstein finally came to the correct equation of the General Relativity in  2015 when he was in Berlin.\footnote{ On November 28 1915 Einstein wrote to Sommerfeld : `` During the past month I had one of the most exciting and strenuous times of my life, but also one of the most successful ones".} In November 1915 Einstein published four papers (every Thursday, November 4, 11, 18 and 25 \cite{E1,E2,E3,E4}). He came to the conclusion that in the left-hand side of the equation (\ref{5})  the Ricci curvature tensor must enter. However, in the first paper   Einstein required invariance of the equation for the gravitation under the unimodular transformations. In the second paper he put even more severe limitation: he required unimodular invariance with $\sqrt{-g}=1$ ($g$ is the determinant of the metric tensor). In the third paper Einstein obtained very important result. From the theory, proposed in the previous paper, for the first time he obtained quantitatively correct value of the angle of the rotation  of the perithelium of Mercury. This gave him confidence of the correctness of the approach he was pursuing .\footnote{Einstein used the field equation in empty space and condition $\sqrt{-g}=1$. In modern calculations the same equation is used in the empty space. The condition $\sqrt{-g}=1$ fixes  the coordinate system.}

Finally, in the November 25 paper Einstein returned back to the requirement of the invariance under general transformations and obtained the correct equation of the General relativity
\begin{equation}\label{8}
R^{\alpha\beta}-\frac{1}{2}g^{\alpha\beta}R =-8\pi G T^{\alpha\beta}.
\end{equation}
 At that time Einstein did not know the Biancci identities (see \cite{Pais}). He assumed the energy-momentum conservation
\begin{equation}\label{9}
T^{\alpha\beta}_{:\rho}=0
\end{equation}
and the term $\frac{1}{2}g^{\alpha\beta}R$ in (\ref{8}) was obtained by Einstein from the requirement (\ref{9}). \footnote{Simultaneously with Einstein the equation (\ref{8}) was obtained by Hilbert  from the variational principle \cite{Hilbert}. There exist a lot of publications in which Einstein and Gilbert papers are compared and discussed (see, for example, \cite{Pais,Todorov:2005rh}). I will cite only the famous Pauli's encyclopedia paper \cite{Pauli}: `` Simultaneously and independently on Einstein covariant equation of the (gravitational) field were established by Gilbert. The Gilbert's presentation was, however,
a little consonant to physicists because, firstly, Gilbert introduced  variational principle axiomatically and, secondly,  and more importantly, his equation was obtained not for arbitrary material system but for special theory of matter proposed by Mie \cite{Mie}}

As we mention before, Einstein  was confused with the equation (\ref{7}) in 1913 when he started to apply the Riemann geometry to the gravitation. Later, after the fundamental equation (\ref{8})
was formulated, Einstein was the first who understood that due to (\ref{7}) it is possible to include in the equation for the metric tensor the cosmological constant $\Lambda$ \cite{E5}. The equation took its final form
\begin{equation}\label{9}
R^{\alpha\beta}-\frac{1}{2}g^{\alpha\beta}R+\Lambda g^{\alpha\beta} =-
8\pi G T^{\alpha\beta}.
\end{equation}


\begin{thebibliography}{99}


\bibitem{Zyla:2020zbs}
  P.~A.~Zyla {\it et al.} [Particle Data Group],
{\em Review of Particle Physics},
  PTEP {\bf 2020} (2020) no.8,  083C01.

\bibitem{Dirac} P.A.M. Dirac, {\em General Theory of Relativity}, Princeton University Press (1996).

\bibitem{Landau}L.D. Landau and E.M. Lifshitz {\em Classical Theory of Fields}, Pergamon Press, Oxford (1975).

\bibitem{Zeldovich}Ya. B. Zeldovich and I.D. Novikov, {\em Relativistic Astrophysics}, Vol. II, Univ. Chicago Press, Chicago (1983)

\bibitem{Pais} Abraham Pais, {\em Subtle is the Lord. The Science and the Life of Albert Einstein} , Clarendon Press, Oxford (1982).

\bibitem{Rev} A. Einstein, Jahrb. Rad. Elektr.  {\bf 4} (1907) 411.



\bibitem{EG} A. Einstein and M Grossmann, {\em Entwurf einer verallgemeinerten Relativitaetstheorie und einer Theorie der Gravitation}   Leipzig: Teubner. (CPAE 4, Doc. 13) (1913).

\bibitem{E1} A. Einstein, PAW (1915) 778.

\bibitem{E2} A. Einstein, PAW (1915) 799.

\bibitem{E3} A. Einstein, PAW (1915) 831.

\bibitem{E4} A. Einstein, PAW (1915) 844.


\bibitem{Hilbert} D. Hilbert, Grundlagen der Physik, 1 Mitt, Goett. Nachr.,(1915) math-nat. p.395.



\bibitem{Todorov:2005rh}
I.~T.~Todorov, {\em Einstein and Hilbert: The Creation of general relativity}, arXiv:physics/0504179.



\bibitem{Pauli} W. Pauli,  Theory of Relativity, New York, Pergamon Press (1958).



\bibitem{Mie} G. Mie, AdP,  {\bf 37} (1912) 511; {\bf 39} (1912) 1; {\bf 40} (1913) 1.

\bibitem{E5} A. Einstein, {\em Cosmological Considerations in the General Theory of Relativity}, Kaeniglich Preussische Akademie der Wissenschaften (1917) 142-152.

\end{thebibliography}
\end{document}